\documentclass[aip,pop,reprint,amsmath,amssymb]{revtex4-1}
\usepackage[dvips]{graphicx}
\usepackage{color}

\newcommand{\comment}[1]{}
\newcommand{\edit}[1]{#1}

\begin{document}
\title{Quasi-stable injection channels in a wakefield accelerator} 
\author{Mara Wiltshire-Turkay}
\author{John P. Farmer}
\author{Alexander Pukhov}
\affiliation{Heinrich-Heine-Universit\"at, 40225 D\"usseldorf, Germany}


\begin{abstract}\noindent
\edit{The influence of initial position on the acceleration of externally-injected electrons in a plasma wakefield is investigated.  Test-particle simulations show previously unobserved complex structure in the parameter space, with quasi-stable injection channels forming for particles injected in narrow regions away from the wake centre.  Particles injected into these channels remain in the wake for a considerable time after dephasing, and as a result achieve significantly higher energy than their neighbours.  The result is relevant to both the planning and optimisation of experiments making use of external injection.}
\end{abstract}

\maketitle

\section{Introduction}
High-energy electrons have a wide range of applications, ranging from fundamental research in particle physics to medical diagnostics and treatment.  However, conventional acceleration methods require long acceleration lengths, which also acts to limit the maximum attainable energy to some tens of GeV.

One promising alternative technique is wakefield acceleration, first proposed by Tajima and Dawson\cite{lwfa-tajimadawson}.  A short driver, either a laser pulse or charged particle beam, is used to excite a plasma wave.  The resulting charge imbalance can lead to large electric fields -- much larger than can be supported by conventional media.  A witness bunch within the wake may then be accelerated.

\edit{The majority of works to date have considered internal injection, due to the relative ease with which high-energy electrons can be achieved.  The first experiments made use of self-injection, in which the plasma wave is driven beyond the wavebreaking limit, allowing plasma electrons to be trapped within the accelerating phase of the wake \cite{lwfa-modena-selfinjection}.  Several schemes have since been proposed to allow better control of the injection process \cite{lwfa-malka-review}, with a view to improve the energy-spread and emittance of the accelerated electron beam.  The use of an external source of electrons may also offer accelerated beams of higher quality \cite{lwfa-grebenyuk-external2014}.}

The AWAKE project\cite{pwfa-AWAKE-2016} will make use of a proton driver, potentially allowing electron acceleration to TeV energies\cite{pwfa-caldwell-protondriven,pwfa-caldwell-selfmodulation}, orders of magnitude higher than current state of the art.  The long proton driver interacts with its own wake via the self-modulation instability, leading to the creation of a train of bunches\cite{pwfa-kumar-selfmodulation}, making efficient wakefield generation possible.  \edit{In the proposed experiment, an externally injected electron bunch will be accelerated.}  Alternatively, high energies could be achieved by making use of a staged laser-driven wakefield accelerator\cite{wakefield-steinke-staged}, with the accelerated electrons from each stage injected into the next.  In both cases the dynamics of external electron injection will play an important role.

In this paper, we investigate the influence of injection position on the energy gain of electrons in a plasma wake.  In Section~\ref{dependence} we show that this dependence is more complex than previously realised, with narrow filaments away from the centre of the wake in which high acceleration can be achieved.  Section~\ref{channels} explains the underlying physics of these quasi-stable injection channels, resulting from electron dephasing with the wake.  The stability of the effect and the applicability to experiments is discussed in Section~\ref{experiment}, and conclusions are drawn in Section~\ref{conclusion}.

\section{Acceleration dependence on injection position}\label{dependence}
\begin{figure}
\includegraphics{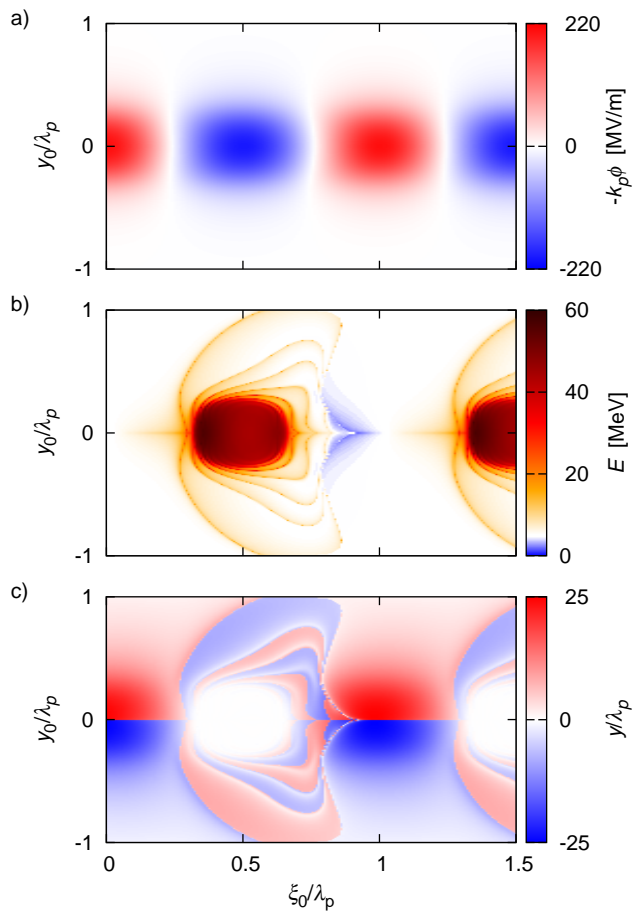}
\caption{Influence of injection position on acceleration after 25~cm ($\sim200\lambda_p$).  a) Wakefield potential, $-k_p\phi$, assumed to be static in this frame.  Blue areas will act to trap electrons.  The driver is located in the $+\xi$ direction.  b) Final electron electron energy against initial position.  c) Final transverse position against initial position.}
\label{fig:scan2d}
\end{figure}

In order to gauge the influence of injection position on energy gain, a parameter scan was carried out using a series of test particle simulations.  A slice in the $x$--$y$ plane was sampled, intersecting with the centre of the wake.  The results are shown in Fig.~\ref{fig:scan2d}.

The electromagnetic wakefields were generated using the fully three-dimensional quasistatic version of the PIC code VLPL \cite{pic-pukhov-quasistatic}.  A section of the associated potential is shown in Fig.~\ref{fig:scan2d}a.  \edit{A short, pre-modulated proton beam was used to give a wake with parameters relevant to the AWAKE experiment, with a driver energy of 400~GeV, and a plasma density of $7\times10^{14}$~cm$^{-3}$ (corresponding to $\lambda_p=2\pi c/\omega_p=1.26$~mm).  For simplicity, injection into a low-amplitude wake, for which the plasma response remains linear, is first considered.  A plasma modulation depth of 8.5\% is therefore used, giving a maximum accelerating field of 220~MV/m.  Comparison to a larger accelerating field, as predicted for AWAKE, as well as a discussion of the use of a laser or electron driver, is made in Section~\ref{experiment}.}

\edit{The wake is assumed to remain static in the co-moving frame, $\xi=x-v_gt$, with $v_g$ the group velocity of the driver, which is valid for a slowly evolving drive beam.  This is somewhat different to the proposed AWAKE experiment, in which the electron beam is injected into the wake during self-modulation, leading to a growing wake with a non-constant phase velocity\cite{pwfa-pukhov-selfmodulationphase,pwfa-schroeder-growthandphasevelocity}.  The case treated here is instead applicable to injection into the developed wake\cite{pwfa-muggli-injection,pwfa-AWAKE-2014}.}

\edit{In the figure, the driver is located at some position in the $+\xi$ direction.  Electrons were placed in the wake with an initial energy of 5~MeV, propagating parallel to the driver.  Experimentally, this could be realised via injection by a co-propagating electron beam entering the plasma behind the drive beam.  Although not considered here, it is worth noting that such injection is complicated by edge-effects upon entering the plasma\cite{pwfa-AWAKE-path,pwfa-lotov-selfmodulationtrapping,pwfa-AWAKE-2014}.}

Particles with different initial positions were propagated in the wakefield using a Boris push.  Calculations show that radiation reaction is negligible in this regime.  The energy attained after a driver propagation distance of 25~cm (corresponding to $\sim200\lambda_p$) is shown in Fig.~\ref{fig:scan2d}b.  As expected, the largest acceleration occurs for particles which are injected into the wake where the fields are both focussing and accelerating.  However, it can immediately be seen that there additionally exist narrow filaments away from the centre of the wake in which high energy can be achieved.

Figure~\ref{fig:scan2d}c shows the final transverse position of injected electrons.  It can be seen that for fixed $\xi$, there exist regions in which electrons are alternately ejected in the positive and negative $y$ directions, corresponding to the particles executing a different number of half-oscillations in the wake before being ejected.  This behaviour is strongly nonlinear, arising due to the nonlinearity of the focussing electric fields, and also due to the larger forward acceleration experienced by particles nearer the axis, which increases both their inertia and the time taken to dephase with the wake. At each transition there is a narrow band in which the final transverse displacement remains small, which correlate with the high-energy filaments.

\section{Quasi-stable injection channels}\label{channels}
\begin{figure}
\includegraphics{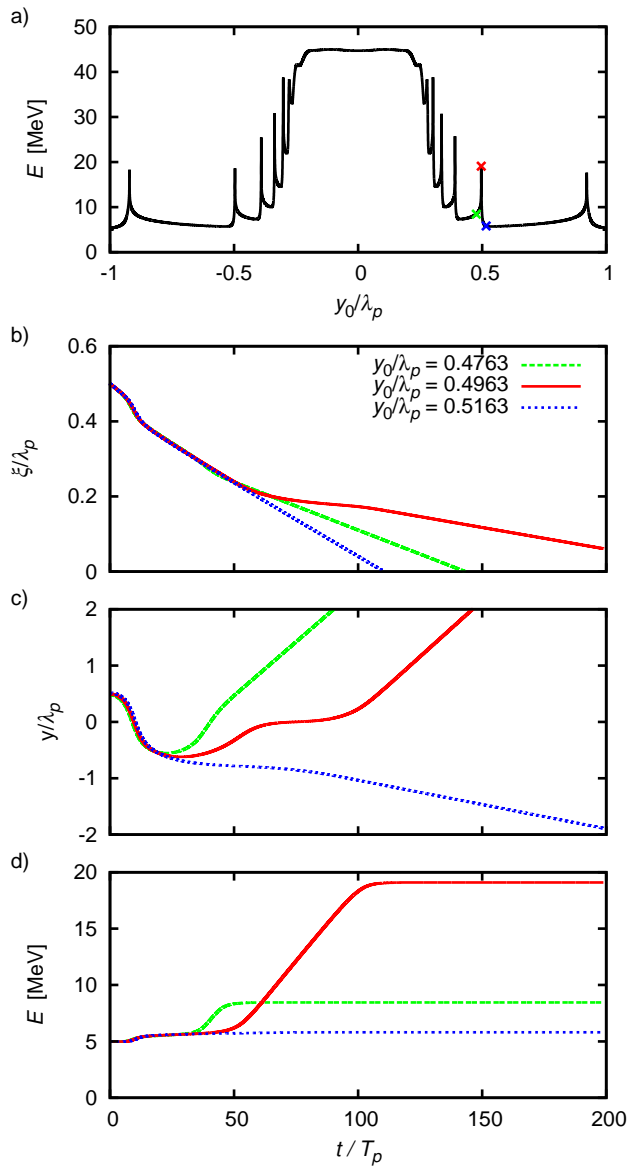}

\caption{Trajectories of individual particles in the wake.  a) Slice showing final electron energy against initial transverse position for $\xi_0=0.5$.  The three initial positions to be tracked are marked.  b) Evolution of $\xi$ against time.  c) Evolution of $y$ against time.  d) Evolution of the energy against time.}
\label{fig:traj}
\end{figure}

In order to better understand the underlying physics of the filament structure, we consider the trajectories of three individual particles.  Figure~\ref{fig:traj}a shows the final energy for varying initial transverse position $y_0$ for a fixed $\xi_0=0.5$ \comment{(equivalent to a slice taken through Fig.~\ref{fig:scan2d})}.  We choose three particles with initial positions, marked on the plot, corresponding to a local maximum in the final energy ($y_0/\lambda_p=0.4963$) and two straddling points, taken at $\pm 0.02\lambda_p$.  Figures ~\ref{fig:traj}b, c and d show the corresponding evolution of $\xi$, $y$ and the energy.

The initial energy of the injected electrons ($5$~MeV) is much lower than that of the driver, and so the particles rapidly fall backwards in the co-moving frame, as seen in Fig~\ref{fig:traj}b.  As they dephase, they pass through the focussing/accelerating phase of the wake and begin to gain energy, as seen in Fig.~\ref{fig:traj}d.

The particles are injected away from the centre of the wake, and so they oscillate in the $y$ plane, and as such experience a smaller average accelerating field than those injected close to the axis.  As a result they do not gain enough energy to become trapped in the wake, and continue to dephase until they reach the defocussing/accelerating phase of the wake.  The defocussing field then causes the electrons to be ejected from the wake in either the $+y$ or $-y$ direction.  The direction in which particles are ejected depends on their position and momentum as they exit the focusing phase, which in turn depends on their initial position.  On the threshold between ejection in the positive and negative directions, there exist particles which achieve a temporary equilibrium, allowing them to remain in the wake for a longer period, and in doing so gain more energy than their neighbours.  For the particles considered here, the central particle remains in the wake more than $50\,T_p$ longer than its neighbours, where $T_p = 2\pi/\omega_p$.  We refer to these trajectories as ``quasi-stable'' -- the particles are globally unstable, but are able to climb the potential gradient in the defocussing phase, gaining more energy as they do.

\begin{figure}
\centering
\includegraphics{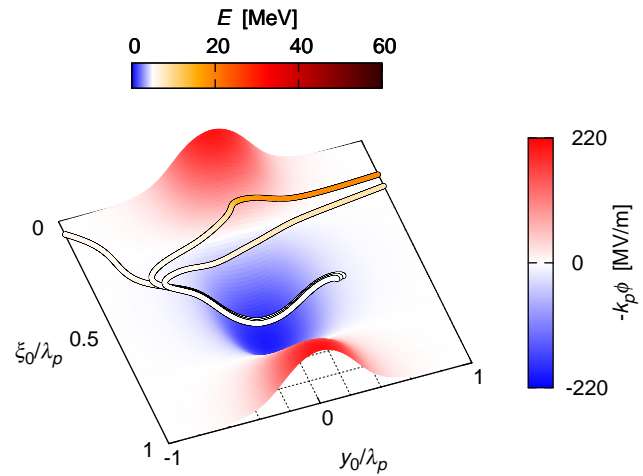}
\caption{Trajectories and energy evolution for the three particles in Fig.~\ref{fig:traj}, superimposed on the wakefield potential.  The potential is shown in the vertical direction, with blue areas acting to trap electrons.  The colour of the electron trajectories corresponds to the energy.}
\label{fig:isotraj}
\end{figure}

The effect is illustrated in Fig.~\ref{fig:isotraj}, which shows the particle trajectories superimposed on the wakefield potential.  Particles oscillate in the trapping potential (blue), before dephasing, falling back to the defocussing phase (red).  Depending on the angle with which the particles approach the defocussing potential, they will be deflected to one side or the other.  The angle of the central particle, however, is such that it climbs the potential gradient in the defocussing phase, rather than immediately being deflected.  The transverse velocity of the particle decreases as it approaches the centre of the wake, allowing it to stay in the accelerating field for significantly longer, leading to increased energies.

These quasi-stable injection channels are a direct consequence of the wakefield structure.  As the accelerating and focussing fields are a quarter-wavelength out of phase, particles can continue to be accelerated after leaving the focussing field.

\section{Stability and applicability to experiments}\label{experiment}
\begin{figure}
\includegraphics{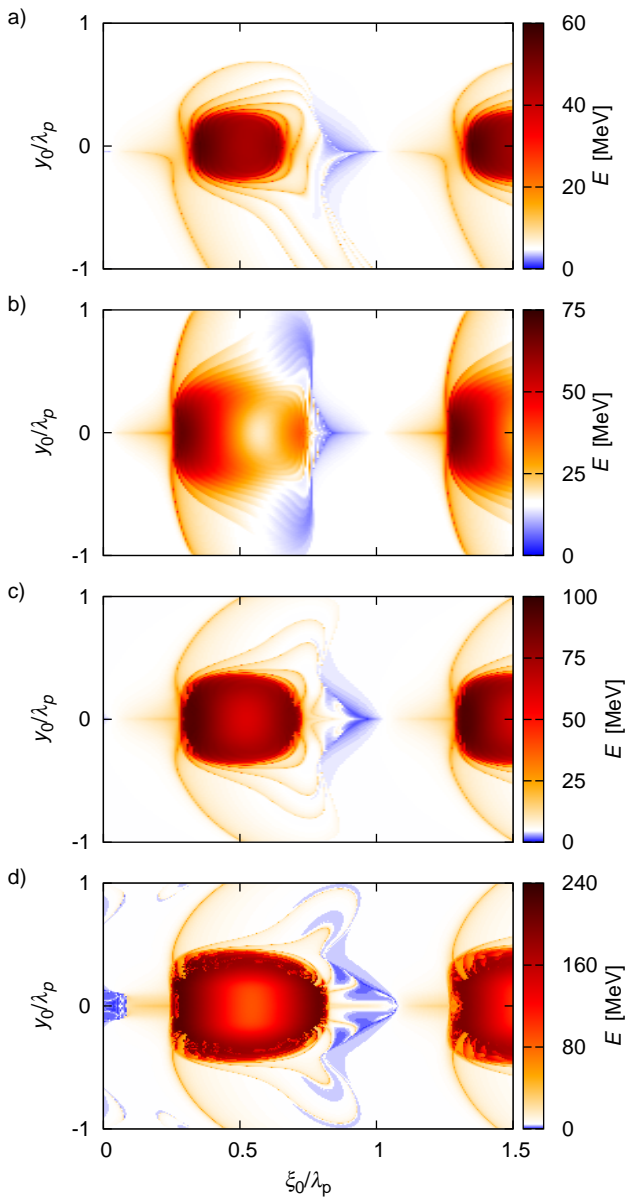}
\caption{Influence of injection position on electron energy after 25~cm ($\sim200\lambda_p$) for different momenta of the injected electrons (Figs.~a,b) and different wake amplitudes (Figs.~c,d).  a) 5~MeV electrons injected from below at a 1$^\circ$ angle to the driver into a 8.5\% modulated wake.  b) 16~MeV electrons injected parallel to the driver into a 8.5\% modulated wake.  c) 5~MeV electrons injected parallel to the driver into a 15\% modulated wake.  d) 5~MeV electrons injected parallel to the driver into a 40\% modulated wake.}
\label{fig:scan2d2}
\end{figure}

The emergence of these quasi-static injection channels appears to be remarkably robust.  Although the wakefields used here are fully electromagnetic, the same structure can be observed using a simple analytical form for a purely electrostatic field.  \edit{Figures \ref{fig:scan2d2}a,b show the influence of changing the initial momentum of the injected electrons.  Other parameters are the same as in Fig.~\ref{fig:scan2d}.  In Fig.~\ref{fig:scan2d2}a, electrons are injected from below at a $1^\circ$ angle, i.e. propgating in the $+\xi$, $+y$ direction.  Although this breaks the symmetry in the $y$ plane, it can be seen that the injection channels remain.  Reflection of the beam, as discussed in Ref.~\onlinecite{pwfa-lotov-optimalangle}, is not observed in this case.  Reflection is expected when the angle relative to the driver is below the optimal angle, which for these parameters is $0.3^\circ$.  Modelling the effect would therefore require a longer propagation distance than the 25~cm used here.}

\edit{The effect of increasing the energy of the injected electrons is shown in Figure~\ref{fig:scan2d2}b.  An electron energy of 16~MeV is used, propagating parallel to the driver.  These faster electrons take longer to dephase, and so execute more oscillations in the trapping potential before being ejected from the wake. As the quasi-stable channels occur on the threshold between different number of oscillations, this results in the channels being more closely spaced in $y_0$.  Changing the velocity of the driver will have a similar effect -- if the driver velocity is decreased, the injected electrons will take longer to dephase.  Conversely, a higher-velocity driver will cause dephasing to occur sooner, and so the injection channels will be more widely spaced.  The main influence of the choice of driver, whether it be an ion, electron or laser beam, is the resulting phase velocity of the wake.  The propagation distance over which the driver can be considered ``slowly-varying'' may also be different.}

\edit{Figures \ref{fig:scan2d2}c,d show the influence modifying the modulation depth of the wakefield.  Again, other parameters are as for Fig.~\ref{fig:scan2d}.  In Fig.~\ref{fig:scan2d2}c, a 15\% modulation depth is used, corresponding to a peak accelerating field of 390~MV/m, near the saturation limit predicted in Ref.~\onlinecite{pwfa-pukhov-selfmodulationphase} for a self-modulating proton beam.  As expected, the peak energy of electrons is larger than those obtained in Fig.~\ref{fig:scan2d} due to the larger accelerating fields.  Although the trapping potential is larger, the transverse momentum of the electrons in the wake is also correspondingly increased, and so little difference is observed in the structure of the injection channels.}

\edit{The modulation depth is further increased to 40\% in Fig~\ref{fig:scan2d2}d.  This is the maximum modulation depth observed in Ref.~\onlinecite{pwfa-lotov-twodimensionalequidistant}, and corresponds to a maximum accelerating field of $\sim1$~GV/m.  It is important to note, however, that in that work, the wake quickly decays after reaching a maximum, while here we assume it is constant.    Again, the larger accelerating field gives rise to an increased peak energy for the accelerated electrons.  For this larger modulation depth, the wake is weakly nonlinear, resulting in a wakefield with anharmonic structure.  The positively charged troughs of the wakefield become larger and shallower than the negatively charged peaks, which leads to a significantly larger trapping area than observed in Figs.~\ref{fig:scan2d} and \ref{fig:scan2d2}c.  However, despite these differences, the injection channels outside the main trapping region remain.}

\begin{figure}
\includegraphics{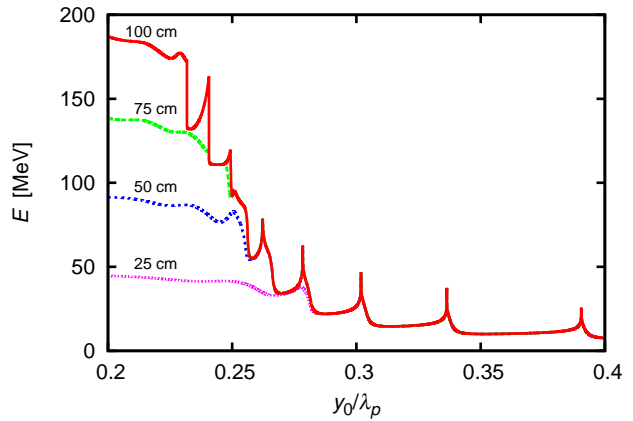}
\caption{Evolution of the energy against injection position over propagation distance.  5~MeV electrons are injected parallel to the driver into an 8.5\% modulated wake.  The energies are shown (from lowest to highest) after 25, 50, 75 and 100~cm.  The energies after 25~cm shown here equivalent to those in Fig.~\ref{fig:traj}a.}
\label{fig:scan1d_distance}
\end{figure}

For longer propagation distances, additional channels form closer to the centre of the wake, as shown in Fig.~\ref{fig:scan1d_distance}.  This is because electrons closer to the wake centre experience a stronger accelerating field, and so take longer to dephase.  As particles ejected at later times will have executed more oscillations in the trapping potential, the injection channels again become more tightly packed.  Over time, the rate at which particles are ejected from the wake decreases, but the effect will occur as long as some particles continue to dephase.

\edit{In this work, we have considered the case of a wakefield with a constant phase velocity.}  As the process relies on electrons dephasing, and ultimately being ejected from the wake, it is not itself useful as an acceleration technique.  However, the result is still relevant to experiments in which external injection is used.  Finding the optimal injection position in such schemes will require careful tuning of the angle and offset of the witness bunch relative to the driver.  The presence of numerous ``false peaks'' in the parameter space topology makes this process more difficult than if there were only a single peak.  Knowledge of this structure is therefore vital to facilitate such tuning.

These quasi-stable channels also allow particles to be accelerated to relatively high energies without becoming fully trapped in the wake.  These electrons will be ejected from the wake throughout the acceleration process, and so precautions must be taken to avoid degradation of any focussing magnets.

\edit{The result may also prove significant in regimes where the phase velocity of the wake is not fixed.  A decrease in the wake velocity could arise, for example, due to erosion of the driver head or the use of a plasma density gradient.  In this case, it may be possible for electrons which have dephased to re-enter the trapping region of the wake.  Such an investigation is beyond the scope of this work, but merits future study.}

\section{Conclusion}\label{conclusion}
The influence of injection position on the energy gain in a wakefield accelerator was investigated through the use of test-particle simulations.  The fields themselves were generated using the quasistatic version of the PIC code VLPL.  We observe, for the first time, the presence of complex structure in the parameter space, with narrow filaments in the injection position, away from the wake axis, in which relatively high energy may be achieved.

It is shown that these filaments correspond to quasi-stable injection channels.  Electrons that are not trapped in the wake dephase, falling back to the defocussing phase of the wakefield.  For a narrow range of initial positions, they approach the defocussing potential with an angle such that they are not immediately ejected from the wake.  This allows them to remain in the wake for significantly longer than their neighbours, and so gain more energy.

The result provides significant insight into the process of external injection, and is relevant for the planning and optimisation of wakefield acceleration experiments in which external injection is used, such as the forthcoming AWAKE project.

\section{Acknowledgements}
JPF and AP would like to acknowledge funding from DFG TR18, EUCard$^2$ EU FP7 and BMBF.  MWT would like to thank G\"otz Lehmann and Matthias Dellweg for helpful discussions.


%

\end{document}